\pdfoutput=1
\documentclass[journal]{IEEEtran}

\usepackage{cite}
\usepackage{graphicx}
\usepackage{algorithm,algorithmic}
\usepackage{ulem}
\usepackage{booktabs}
\usepackage{amsmath}
\usepackage{amssymb}
\usepackage{booktabs}
\usepackage{stackengine}
\usepackage{mathrsfs}
\DeclareMathAlphabet{\mathcal}{OMS}{cmsy}{m}{n}
\usepackage{newtxtext}
\DeclareSymbolFont{matha}{OML}{txmi}{m}{it}% txfonts
\usepackage{array}
\usepackage{graphicx}
\usepackage{subcaption}
\usepackage{caption}
\usepackage{url}
\usepackage{hyperref}
\usepackage{color}
\usepackage{graphicx}
\usepackage{fancyhdr}
\usepackage{lipsum}
\fancypagestyle{plain}{\pagestyle{fancy}}

\begin{document}

\pagestyle{fancy}
\fancyhead[C]{\fontsize{6}{8} \fontfamily{phv}\selectfont This article has been accepted for publication in IEEE Transactions on Power Systems. This is the author's version which has not been fully edited and content may change prior to final publication. Citation information: DOI 10.1109/TPWRS.2022.3142105}
\fancyfoot[C]{\fontsize{6}{6} \fontfamily{phv}\selectfont © 2022 IEEE. Personal use is permitted, but republication/redistribution requires IEEE permission. See https://www.ieee.org/publications/rights/index.html for more information.}
\fancyhead[R]{\thepage}

\title{Bi-level Volt/VAR Optimization in Distribution Networks with Smart PV Inverters}

\author{Yao Long,~\IEEEmembership{Student Member,~IEEE,}
        Daniel S. Kirschen,~\IEEEmembership{Fellow,~IEEE}% <-this % stops a space
\thanks{Y. Long and D. S. Kirschen are with the Department of Electrical and Computer Engineering, University of Washington, Seattle, WA, 98195 USA (e-mail: longyao@uw.edu; kirschen@uw.edu). \textit{(Corresponding author: Yao Long)}}}

\newcommand\ubar[1]{\stackunder[1.2pt]{$#1$}{\rule{.8ex}{.075ex}}}
%\markboth{}%

\maketitle \thispagestyle{fancy} 
\begin{abstract}
Optimal Volt/VAR control (VVC) in distribution networks relies on an effective coordination between the conventional utility-owned mechanical devices and the smart residential photovoltaic (PV) inverters. Typically, a central controller carries out a periodic optimization and sends setpoints to the local controller of each device. However, instead of tracking centrally dispatched setpoints, smart PV inverters can cooperate on a much faster timescale to reach optimality within a PV inverter group. To accommodate such PV inverter groups in the VVC architecture, this paper proposes a bi-level optimization framework. The upper-level determines the setpoints of the mechanical devices to minimize the network active power losses, while the lower-level represents the coordinated actions that the inverters take for their own objectives. The interactions between these two levels are captured in the bi-level optimization, which is solved using the Karush-Kuhn-Tucker (KKT) conditions. This framework fully exploits the capabilities of the different types of voltage regulation devices and enables them to cooperatively optimize their goals. Case studies on typical distribution networks with field-recorded data demonstrate the effectiveness and advantages of the proposed approach.
\end{abstract}

\begin{IEEEkeywords}
Bi-level optimization, distribution network, photovoltaic system, smart inverter, Volt/VAR control.
\end{IEEEkeywords}

\section*{Nomenclature}
\addcontentsline{toc}{section}{Nomenclature}

\subsection{Acronyms}
\begin{IEEEdescription}[\IEEEusemathlabelsep\IEEEsetlabelwidth{$V_1,V_2,V_3$}]
\item[ADMS] Advanced Distribution Management System
\item[CB] Capacitor Bank
\item[OLTC] On-Load Tap Changer
\item[PV] Photovoltaic
\item[VVC] Volt/VAR Control
\end{IEEEdescription}

\subsection{Sets and Indices}
\begin{IEEEdescription}[\IEEEusemathlabelsep\IEEEsetlabelwidth{$V_1,V_2,V_3$}]
\item[$\mathcal{E}$, ${ij}$] Set, index of all the network branches
\item[$\mathcal{N}$, $i$] Set, index of all the network nodes
\item[$\mathcal{N_G}$] Set of nodes with PV
\item[$k$] Index of the iteration cycles in the lower-level
\item[$t$] Index of the dispatch periods in the upper-level

\end{IEEEdescription}
\subsection{Variables}
\begin{IEEEdescription}[\IEEEusemathlabelsep\IEEEsetlabelwidth{$V_1,V_2,V_3$}]
\item[$l_{ij}$] Squared current magnitude on branch $ij$
\item[$n_{i}^{C}$] Number of the connected CB units at node $i$
\item[$n_{ij}^{T}$] Tap position of the OLTC on branch $ij$
\item[$q_i^{g}$] Reactive power output of the PV at node $i$
\item[$v_i$]Squared voltage magnitude at node $i$
\item[$P_{ij}$, $Q_{ij}$] Active, reactive power flow on branch $ij$

\end{IEEEdescription}
\subsection{Parameters}
\begin{IEEEdescription}[\IEEEusemathlabelsep\IEEEsetlabelwidth{$V_1,V_2,V_3$}]

\item[$\bar{l}_{ij}$] Upper limit for the squared current magnitude on branch $ij$
\item[$\hat{p}_i^g$, $p_i^g$] Forecast, actual active power output of the PV at node $i$
\item[$\hat{p}_i^{l}$, $\hat{q}_i^{l}$] Forecast active, reactive load at node $i$
\item[$\bar{q}_i^g$, $\ubar{q}_i^g$] Upper, lower limit of the reactive power output of the PV at node $i$
\item[$r_{ij}$, $x_{ij}$] Resistance, reactance of branch $ij$
\item[$s_i^g$] Capacity of the PV inverter at node $i$
\item[$\bar{v}_i$, $\ubar{v}_i$] Upper, lower limit for the squared voltage magnitude at node $i$
\item[$\Delta \bar{n}_{i}^{C}$] Maximum change of the CB units at node $i$ between two dispatch periods
\item[$\Delta \bar{n}_{ij}^{T}$] Maximum tap change of the OLTC on branch $ij$ between two dispatch periods
\item[$N_{i}^{C}$] Total number of the CB units at node $i$
\item[$N_{ij}^{T}$] Total tap positions of the OLTC on branch $ij$
\item[$Q_{i}^{C}$] Total capacity of the CB units at node $i$

\end{IEEEdescription}
\IEEEpeerreviewmaketitle

\section{Introduction}
\IEEEPARstart{v}{olt}/VAR control (VVC) manages voltage levels and reactive power in  distribution networks to reduce active power losses and maintain acceptable voltage magnitudes using on-load tap changers (OLTCs) and capacitor banks (CBs) \cite{VVC}. However, a high penetration of residential rooftop photovoltaic (PV) generation can cause rapid voltage fluctuations that these conventional mechanical devices are not designed to handle \cite{Traditional1, Traditional2}. On the other hand, smart PV inverters with sensing, communication and computing capabilities are able to act autonomously to provide fast and flexible reactive power compensation for voltage regulation \cite{Smart}. They are thus encouraged to participate in VVC by recent amendments of the IEEE 1547 Standard\cite{1547}.

This paper develops a bi-level optimization framework for the VVC in distribution networks. The upper-level optimization determines the periodic dispatch of the slower mechanical devices to minimize the network active power losses. It is formulated as a mixed integer second order cone programming (MISOCP) model that contains a nested lower-level optimization as the constraint on PV inverter reactive output. This optimization problem is intended to be centrally solved in an Advanced Distribution Management System (ADMS) based on the network information after replacing the lower-level with its Karush-Kuhn-Tucker (KKT) conditions. The lower-level optimization takes the settings of OLTCs and CBs as inputs and models the coordinated outputs of the smart PV inverters. These inverters remove sudden voltage limit violations between upper-level dispatch periods while optimizing their group objectives, e.g. minimizing the reactive power cost or equitably sharing the reactive power contributions for voltage regulation. The lower-level optimization is solved in a distributed manner as the PV inverters adjust their outputs autonomously based on real-time voltage measurements and neighboring communication. 

\subsection{Literature Review}
Various strategies have been proposed for the coordinated voltage regulation in distribution networks. Based on the autonomy level of the inverters, they can be broadly classified into three categories.

Strategies of the first category periodically optimize the setpoints and send them to the local controller of each device, which implements these setpoints directly. The autonomous VVC control capability of the smart inverters is thus not fully exploited. For example, Jha et al. \cite{Twolevel} solve a mixed integer linear programming (MILP) problem to set the CBs and OLTCs, while the inverters are dispatched based on the solution of a nonlinear programming (NLP) problem that more accurately models the power flows. To accelerate the optimization and maintain customer data privacy, the modified alternating direction method of multipliers (ADMM), which can handle discrete variables, is used in \cite{ADMM1, ADMM2} to solve the centralized optimization problem with bus-level control agents. Two-stage stochastic optimization \cite{Stochastic} and robust optimization \cite{Robust} algorithms have been implemented to handle uncertainty on the renewable generation in the optimal reactive power dispatch. These control strategies are open-loop and implement feedforward optimization techniques that rely on network information and forecasts. They are generally less robust to errors in the model and the forecasts and also less effective at dealing with fast voltage fluctuations \cite{ReviewD}.

In the second category of strategies, the smart PV inverters follow the centrally-optimized setpoints while adjusting their reactive outputs based on real-time local measurements to improve the control performance. In \cite{Droop1}, the inverter generates the reactive power dispatched by the central controller when its local voltage is within the allowable operational limits, otherwise the droop control is activated. Malekpour et al. further added a distributed algorithm to allocate the voltage regulation burden among inverters in proportion to their capacities \cite{Consensus1}. Besides the reactive power setpoint, inverters can also be controlled to track the voltage reference \cite{Droop2}. Instead of just considering the upper and lower reactive power limits of the inverters, several authors \cite{Combine1, Combine2, Combine3} model their reactive power generation as a function of the local voltage in the optimization model, which allows the central controller to also optimize the local droop control parameters for the PV inverter. These multi-level control strategies combine the system-wide feedforward optimization with the real-time local feedback control to increase their robustness to forecast errors and fast fluctuations in renewable generation. However, the regulation capacity of the PV inverters is not effectively coordinated in real-time. Instantaneous voltage violations can happen when the linear droop control in \cite{Droop1, Droop2, Combine1, Combine2, Combine3} lacks locally available reactive power~\cite{Communication}. Smart inverters in \cite{Consensus1} coordinate their reactive outputs based on consensus algorithm. However, this coordination is not considered by the central controller. Large deviations between the actual inverter output and their centrally optimized setpoint might occur, which affects the optimality of the VVC strategy.

The third category of strategies emerged recently with the development of smart inverter technology. These inverters continuously and autonomously adjust their output while cooperating with each other through neighboring communication. Strategies in \cite{Consensus2, Consensus3} implement a leader-follower consensus algorithm, which enables the smart inverters to regulate the network voltage magnitude while sharing the regulation burden according to their maximum available capacity. Other strategies rely on distributed optimization, such as dual ascent algorithm\cite{Communication, Dualascent0, Dualascent1, Dualascent2} and primal-dual gradient algorithm \cite{Primaldual1}. By emulating the iterative steps of these algorithms, PV inverter groups achieve optimal coordination in real-time to remove local voltage violations while minimizing power losses \cite{Communication, Dualascent0, Dualascent1}, reactive power cost \cite{Dualascent2} or a weighted sum of both \cite{Primaldual1}. These strategies fully exploit the flexibility and fast response capability of smart PV inverters. Moreover, they do not need to model the variations of load or renewable generation explicitly, instead, the smart PV inverters adjust their output by constantly monitoring the dynamic system. Such feedback optimization-based strategies are more robust to modeling errors and uncertainty on the renewable generation. They can also improve the dynamic performance of the closed-loop system \cite{ReviewD}. However, the slower mechanical devices with discrete control actions are unable to participate in such strategies and their setpoints are still set by the central controller. The coordination of the mechanical devices with the cooperative smart inverter group has not yet been studied.

\subsection{Paper Contribution and Organization}
As the PV inverters gain more autonomy, they are able to take more sophisticated real-time cooperative control actions to optimize their own VVC objectives, which might even differ from that of the utility's. Hence, it becomes less valid to model their operation characteristics as purely following upper-level setpoints or executing droop control based solely on their local voltages. On the other hand, as the PV inverters are playing a more active and important role in the distribution network voltage regulation, their interactions with the centrally controlled mechanical devices need to be handled more carefully. Otherwise, unnecessary control actions or even operational conflicts might occur \cite{Interaction}. A new framework is thus needed to effectively incorporate the smart PV inverters in VVC.

The main contributions of this paper are:
\begin{itemize}
    \item A bi-level optimization model which captures the interactions between the utility and consumer-owned devices. The upper-level contains the lower-level as a constraint, which reflects the influence of the residential PV inverters' reactive output on the network active power losses. On the other hand, the lower-level takes the OLTC and CB settings determined by the upper-level as parameters, as those values in turn affect the local voltage magnitude of the PV inverters. 
    \item A solution based on the KKT conditions that guarantees an optimal control of both the mechanical devices and the PV inverters. The upper-level becomes a solvable MISOCP model after replacing the lower-level by its KKT conditions. The ADMS solves this problem to optimally schedule the mechanical devices while taking into account the PV inverters' response. The smart PV inverters emulate the KKT condition-based distributed optimization algorithms to adjust their reactive output with local voltage measurement, which iteratively solves the lower-level optimization under the upper-level determined OLTC and CB settings.
    \item A two-timescale VVC framework that combines the advantages of feedforward and feedback optimizations. The upper-level deals with the slower voltage variations caused by regular daily changes in load and PV generation. It makes use of the network information available in the ADMS to achieve a system-wide optimal coordination. In real-time, the fast and autonomous PV inverters cooperate to remove the instantaneous voltage violations, which improves the dynamic voltage control performance and the robustness of the framework. 
\end{itemize}

The remainder of this paper is organized as follows. Section II describes the details of the bi-level optimization model. Section III presents the solution approach and its implementation. Section IV demonstrates and analyzes simulation results. Section V concludes.

\section{Bi-level Optimization Model}
This section describes the objectives and constraints of the upper and lower-level optimization models. The composite bi-level optimization model is then presented.

\subsection{Upper-level}
The upper-level dispatches OLTCs and CBs to minimize the network active power losses for the dispatch period $t$:
\begin{equation} \label{eq: obj}
\begin{split}
     F = \sum_{(i, j) \in \mathcal{E}}r_{ij}l_{ij, t}.
\end{split}
\end{equation}
Meanwhile, it must satisfy the following constraints ${\forall i\in \mathcal{N}}$, ${\forall (i, j) \in \mathcal{E}}$.
\subsubsection{System Operational Constraints}
The DistFlow model proposed in \cite{DistFlow} has been extensively used for modeling distribution networks because of its accuracy and computational efficiency. It formulates the power flow constraints as follows: 
\begin{equation} \label{eq: DF1}
P_{ij, t} = \sum_{k:(j,k) \in \mathcal{E}}P_{jk, t} + r_{ij}l_{ij, t} + \hat{p}_{j,t}^l - \hat{p}_{j,t}^g
\end{equation}
\begin{equation} \label{eq: DF2}
Q_{ij, t} = \sum_{k:(j,k) \in \mathcal{E}}Q_{jk, t} + x_{ij}l_{ij, t} + \hat{q}_{j,t}^l - q_{j,t}^g - q_{j,t}^c
\end{equation}
\begin{equation} \label{eq: DF3}
v_{j,t} = v_{i,t} - 2(r_{ij}P_{ij,t}+x_{ij}Q_{ij,t})+(r_{ij}^2+x_{ij}^2)l_{ij,t}
\end{equation}
\begin{equation} \label{eq: nonconvex}
l_{ij,t}v_{i,t}=P_{ij,t}^2+Q_{ij, t}^2
\end{equation}
Since the equality constraint (\ref{eq: nonconvex}) is non-convex, we relax it to an inequality constraint using  the method proposed in \cite{SOCP} to obtain a global optimal solution:
\begin{equation} \label{eq: DF4}
\Big\|\begin{bmatrix}
2P_{ij, t} & 2Q_{ij,t} & l_{ij,t}-v_{i,t}
\end{bmatrix}^T \Big\|_2 \le l_{ij,t} + v_{i,t}.
\end{equation}
On the primary and secondary sides of an OLTC, (\ref{eq: DF3}) is modified as follows:
\begin{equation} \label{eq: OLTC}
\frac{v_{j,t}}{T_{ij,t}} = v_{i,t} - 2(r_{ij}P_{ij,t}+x_{ij}Q_{ij,t})+(r_{ij}^2+x_{ij}^2)l_{ij,t}
\end{equation}
where ${T_{ij}}$ is the squared turns ratio. For each tap position ${n \in \{1, ..., N_{ij}^{T}\}}$, there is a corresponding $T_{ij}^n$. Thus,
\begin{equation}
\frac{1}{T_{ij,t}} = \sum_{n=1}^{N_{ij}^{T}} \frac{w_{ij,t}^n}{T_{ij}^n}
\end{equation}
where
\begin{equation} \label{eq: ST1}
\sum_{n=1}^{N_{ij}^{T}} w_{ij,t}^n = 1, \quad w_{ij,t}^n \in \{0, 1\},
\end{equation}
i.e. only one of the auxiliary binary variables $w_{ij,t}^n$ is nonzero, which specifies the selection of one turns ratio. Moreover, the nodal voltages and branch currents must be maintained within the acceptable range:
\begin{equation} \label{eq: S1}
\ubar{v}_i \le v_{i,t} \le \bar{v}_{i},
\end{equation}
\begin{equation} \label{eq: S2}
l_{ij,t} \le \bar{l}_{ij}.
\end{equation}

\subsubsection{Control Devices Operational Constraints}
The tap change between two dispatch periods should be limited to reduce the mechanical wear and tear:
\begin{equation} \label{eq: OLTC1}
|n_{ij,t}^{T} - n_{ij,t-1}^{T}| \le \Delta \bar{n}_{ij}^{T}
\end{equation}
\begin{equation} \label{eq: OLTC3}
n_{ij,t}^{T} \in \{1, ..., N_{ij}^{T}\}.
\end{equation}
Similarly, the operational constraints for CBs are:
\begin{equation} \label{eq: CB1}
|n_{i,t}^{C} - n_{i,t-1}^{C}| \le \Delta
\bar{n}_{i}^{C}
\end{equation}
\begin{equation} \label{eq: CB3}
n_{i,t}^{C} \in \{0, ..., N_{i}^{C}\}.
\end{equation}
The reactive power injection from CBs is
\begin{equation} \label{eq: CB4}
q_{i,t}^c = \frac{n_{i,t}^{C}}{N_{i}^{C}}Q_{i}^{C}.
\end{equation}
The equality constraint on the PV inverter reactive output is a nested optimization problem corresponding to their coordinated autonomous operation:
\begin{equation} \label{eq: PV}
\textbf{q}_t^g \in \arg \min_{\textbf{q}_t^g} \{f(\textbf{q}_t^g): \textbf{G} \le 0\}
\end{equation}
where ${\textbf{q}_t^g}$ is the vector containing ${q_{i,t}^g}$, ${\forall i \in \mathcal{N_G}}$. $f(\textbf{q}_t^g)$ and $\textbf{G}$ are the objective function and set of constraints of the lower-level optimization, which are detailed in the next subsection. 

\subsection{Lower-level}
In the lower-level, the PV inverters cooperate to achieve an optimal utilization of their reactive power while removing instantaneous voltage limit violations:
\begin{equation} \label{eq: lower level}
\begin{split}
\min_{\textbf{q}_t^g}  \quad\quad& f(\textbf{q}_t^g) \\
\text{s.t.} \quad\quad&\ubar{v} \le v_{i,t} \le \bar{v}, \: \forall i \in \mathcal{N_G} \\
&\ubar{q}_{i,t}  \le q_{i,t}^g \le \bar{q}_{i,t}, \: \forall i \in \mathcal{N_G}
\end{split}
\end{equation}
where ${\bar{q}_{i,t}}$ and ${\ubar{q}_{i,t}}$ are the upper and lower PV inverter reactive output limits, ${\bar{q}_{i,t} = \sqrt{(s_i^g)^2 - (p_{i,t}^g)^2}}$ and ${\ubar{q}_{i,t} = -\bar{q}_{i,t}}$. Different objective functions can be set:
\subsubsection{Minimize Reactive Power Cost}
The provision of reactive power for VVC normally requires excess capacity on the inverters, and thus over-sizing hardware design or active power curtailment. The PV inverters could seek to minimize their reactive power cost \cite{Dualascent2, Primaldual1}:
\begin{equation} \label{eq: cost}
f(\textbf{q}_t^g) =  \sum_{i \in \mathcal{N_G}} a_i^2(q_{i,t}^g)^2+b_i(q_{i,t}^g) + c_i
\end{equation}
where $a_i$, $b_i$, $c_i$ are synthetic cost parameters.
\subsubsection{Minimize Network Active Power Losses}
The PV inverter group could also minimize the network active power losses associated with their reactive output. For ease of implementation, the objective function is usually defined as \cite{Communication, Dualascent0}:
\begin{equation} \label{eq: loss}
f(\textbf{q}_t^g) =  ({\textbf{q}_t^g})^TX_{gg}(\textbf{q}_t^g)
\end{equation}
where ${X_{gg}}$ is a submatrix of $X$, whose elements are:
\begin{equation} \label{eq: Xgg}
X_{ij} := 2\sum_{(h,k)\in \mathcal{P}_{i} \cap \mathcal{P}_{j} } x_{hk}
\end{equation}
where $\mathcal{P}_{i}$ is the set of distribution lines on the unique path from the substation to bus $i$. Partition $X$ based on ${i \in \mathcal{N_G}}$ or ${i \notin \mathcal{N_G}}$:
\begin{equation} \label{eq: Xg}
X = \begin{bmatrix}
X_{gg} & X_{gl} \\
X_{lg} & X_{ll} 
\end{bmatrix}.
\end{equation}
${X_{gg}}$ is the submatrix associated with the PV nodes and it is positive definite \cite{Communication,Primaldual1}. Besides, this objective function also promotes uniform voltage drops \cite{Communication}.
\subsubsection{Equalize Utilization Ratios}
The utilization ratio of an inverter in VVC is the proportion of its maximum available reactive power capacity used for voltage regulation, i.e., ${u_i = q_i^g / \bar{q}_i}$. To prevent excessive use and early saturation of certain inverters, the PV inverter group can cooperate to operate at the same utilization ratio. The leader inverter, typically the inverter at the end node of the feeder because it usually experiences the largest voltage deviation, measures its local voltage to determine this ratio \cite{Consensus2,Consensus3}. In this case, the lower-level optimization objective can be formulated as:
\begin{equation} \label{eq: cons}
f(\textbf{q}_t^g) =  \sum_{i \in \mathcal{N_G}} \frac{X_{li}}{2\bar{q}_{i,t}}(q_{i,t}^g)^2
\end{equation}
where ${l\in \mathcal{N_G}}$ is the index of the leader inverter. The proof is detailed in the Appendix. 

Moreover, $f(\textbf{q}_t^g)$ can also be defined as a weighted sum of these objectives. For example, the PV inverter objective function in \cite{Primaldual1} combines (\ref{eq: cost}) and (\ref{eq: loss}). For a network that contains several groups of PV inverters located in different control zones and where there is no inter-group coordination, the lower-level can be flexibly modified into a set of two or more optimization models. Each of these models describes the coordinated action of a PV inverter group. 

\subsection{Bi-level Optimization Model}
The bi-level optimization model is as follows:
\begin{equation} \label{eq: nest}
\begin{split}
\min_{n_{ij,t}^{T}, \: n_{i,t}^{C}} \quad& F\\
\text{s.t.} \quad&(\ref{eq: DF1})-(\ref{eq: DF3}), \, (\ref{eq: DF4})-(\ref{eq: CB4}),\\
&\textbf{q}_t^g \in \arg \min_{
\textbf{q}_t^g}  \Big\{f(\textbf{q}_t^g):\ubar{v} \le v_{i,t} \le \bar{v}, \\ &\quad\quad\quad\quad\quad\quad\;\;\ubar{q}_{i,t} \le q_{i,t}^g \le \bar{q}_{i,t}, \; \forall i \in \mathcal{N_G} \Big\}
\end{split}
\end{equation}
In this model, each of the two optimization tasks has its own objective, constraints and decision variables. On the other hand, these two tasks are coupled, as the active network losses depend on the PV inverter reactive output and the PV inverters' local voltage magnitude is also influenced by the settings of the mechanical devices.

\subsection{Discussion}
Traditional single-level optimization models generally treat $\textbf{q}_t^g$ as the upper-level decision variable. This is based on the assumption that the PV inverters rely on the upper-level to coordinate their outputs. They either follow the dispatched setpoints directly \cite{Twolevel, ADMM1, ADMM2, Stochastic, Robust} or execute local droop control with upper-level optimized parameters \cite{Combine1, Combine2, Combine3}. However, when the PV inverters implement strategies as proposed in \cite{Consensus2, Consensus3, Communication, Dualascent0, Dualascent1, Dualascent2, Primaldual1} to optimize their outputs autonomously, their reactive generation are not directly controllable by the upper-level and these single-level optimization models become less valid. In this case, the proposed bi-level extension is necessary. This bi-level optimization model captures the operation characteristics of the coordinated PV inverter group and thus allows the mechanical devices to be optimally scheduled while considering the response of the smart PV inverters. Therefore, an effective inter-level coordination can be established for system-wide optimal VVC.

\section{Solution Approach and Implementation}
This section describes the solution approach and the implementation of the bi-level optimization model. The key idea involves representing the lower-level optimization by its KKT conditions. The ADMS and the smart PV inverters operate to ensure these conditions are satisfied in their own ways.

\begin{figure*}[ht!]
\centering
\includegraphics[width=6.75in]{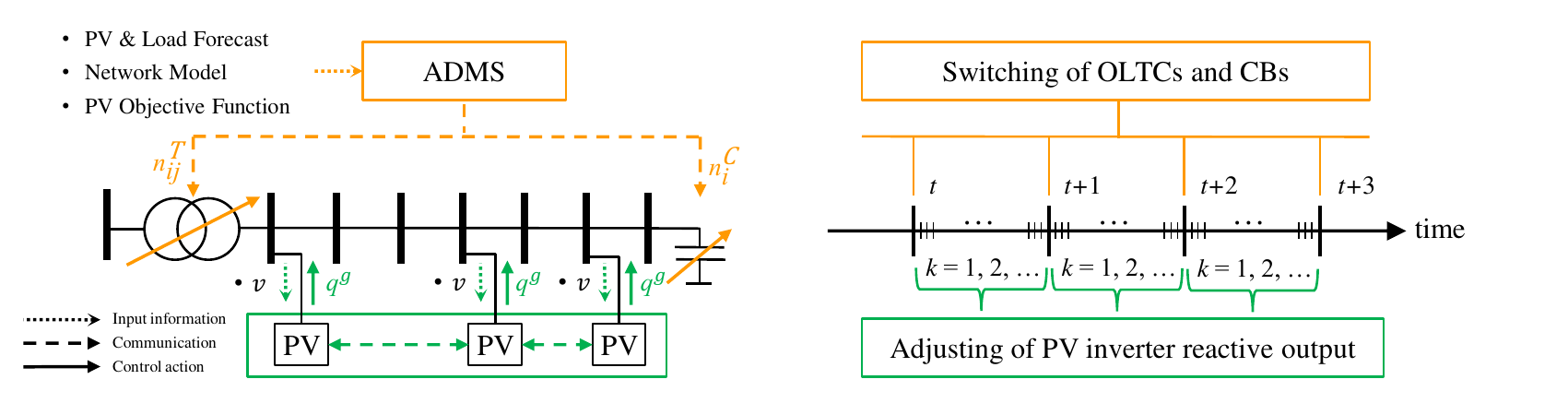}
\caption{Overview of the implementation of the bi-level optimization model.}
\label{Framework}
\end{figure*}

\subsection{KKT Conditions}
With objective function (\ref{eq: cost}), (\ref{eq: loss}) or (\ref{eq: cons}), the lower-level optimization problem~(\ref{eq: lower level}) is convex. Assume the Slater condition is satisfied, i.e. there exist ${\ubar{q}_i < q_{i,t}^g < \bar{q}_i}$, ${\forall{i \in \mathcal{N_G}}}$ such that ${\ubar{v} < v_{i,t} < \bar{v}}$, ${\forall{i \in \mathcal{N_G}}}$, then this problem can be replaced by its KKT conditions:

\begin{equation} \label{eq: lagrange}
\begin{split}
&\nabla_{q_{i,t}^g}\bigg\{f(\textbf{q}_t^g)+\sum_{i \in \mathcal{N_G}}\ubar{\lambda}_{i,t}(\ubar{v} - v_{i,t}) + \sum_{i \in \mathcal{N_G}}\bar{\lambda}_{i,t}(v_{i,t}-\bar{v}) \\ 
&\quad \quad +\sum_{i \in \mathcal{N_G}}\ubar{\mu}_{i,t}(\ubar{q}_{i,t} - q_{i,t}^g) + \sum_{i \in \mathcal{N_G}}\bar{\mu}_{i,t}(q_{i,t}^g-\bar{q}_{i,t})\bigg\} = 0 \\
\end{split}
\end{equation}
\begin{equation} \label{eq: vq}
\ubar{v} \le v_{i,t} \le \bar{v}, \quad \ubar{q}_{i,t}  \le q_{i,t}^g \le \bar{q}_{i,t}
\end{equation}
\begin{equation} \label{eq: lambda}
\ubar{\lambda}_{i,t}(\ubar{v} - v_{i,t}) = 0, \quad \bar{\lambda}_{i,t}(v_{i,t}-\bar{v}) = 0
\end{equation}
\begin{equation} \label{eq: mu}
\ubar{\mu}_{i,t}(\ubar{q}_{i,t} - q_{i,t}^g) = 0, \quad \bar{\mu}_{i,t}(q_{i,t}^g-\bar{q}_{i,t}) = 0
\end{equation}
\begin{equation} \label{eq: dual}
\ubar{\lambda}_{i,t}, \: \bar{\lambda}_{i,t}, \: \ubar{\mu}_{i,t}, \: \bar{\mu}_{i,t} \geq 0
\end{equation}
${\forall i \in \mathcal{N_G}}$. In which, $\ubar{\lambda}_{i,t}$, $\bar{\lambda}_{i,t}$, $\ubar{\mu}_{i,t}$, $\bar{\mu}_{i,t}$ are dual variables.

\subsection{Control of OLTCs and CBs}
As Fig.~\ref{Framework} shows, The ADMS dispatches OLTCs and CBs based on a full knowledge of the network, which includes forecast of the load and PV generation, network model, as well as the objective function selected by the PV owners. For each dispatch period, it constructs the bi-level optimization model described in Section II and then uses the single-level reduction approach \cite{Bilevel} to replace the lower-level with its KKT conditions. The sensitivity of the squared voltage to the inverter reactive output $\partial{v_{i,t}}/\partial{q_{j,t}^g}$ in (\ref{eq: lagrange}) can be approximated by $X_{ij}$ \cite{Validation} or calculated based on the Jacobian matrix. The bi-linear constraint stemming from the complementary slackness condition can be handled using the big-M method, which replaces Eqs. (\ref{eq: lambda})-(\ref{eq: mu}) by:
\begin{equation} \label{eq: vl}
\ubar{\lambda}_{i,t} \le M\ubar{\sigma}_{i,t}, \quad 
v_{i,t}-\ubar{v} \le M(1-\ubar{\sigma}_{i,t})
\end{equation}
\begin{equation} \label{eq: vu}
\bar{\lambda}_{i,t} \le M\bar{\sigma}_{i,t}, \quad 
\bar{v} - v_{i,t}\le M(1-\bar{\sigma}_{i,t})
\end{equation}
\begin{equation} \label{eq: ql}
\ubar{\mu}_{i,t} \le M\ubar{\delta}_{i,t}, \quad 
q_{i,t}^g-\ubar{q}_{i,t} \le M(1-\ubar{\delta}_{i,t})
\end{equation}
\begin{equation} \label{eq: qu}
\bar{\mu}_{i,t} \le M\bar{\delta}_{i,t}, \quad 
\bar{q}_{i,t} - q_{i,t}^g\le M(1-\bar{\delta}_{i,t})
\end{equation}
${\forall i \in \mathcal{N_G}}$, where $M$ is a large number, $\ubar{\sigma}_{i,t}$, $\bar{\sigma}_{i,t}$, $\ubar{\delta}_{i,t}$, $\bar{\delta}_{i,t}$ are auxiliary binary variables. Consequently, the bi-level optimization problem becomes a solvable single-level MISOCP problem:
\begin{equation} \label{eq: fo}
\begin{split}
\min_{n_{ij,t}^{T}, \: n_{i,t}^{C}, \: \textbf{q}_t^g, \: \underline{\lambda}_{i,t}, \: \bar{\lambda}_{i,t}, \: \underline{\mu}_{i,t}, \: \bar{\mu}_{i,t}} \quad &\quad\quad F \\
\text{s.t.} \quad  (\ref{eq: DF1})-(\ref{eq: DF3}), \: (\ref{eq: DF4})-(\ref{eq: CB4}), \: &(\ref{eq: lagrange}), \: (\ref{eq: vq}), \: (\ref{eq: dual})-(\ref{eq: qu}).\\
\end{split}
\end{equation}
The ADMS solves this problem using off-the-shelf optimization tools. At each dispatch period $t$, it delivers the setpoints to the local controllers of OLTCs and CBs via the supervisory control and data acquisition (SCADA) system or the communication link between the devices and ADMS. These setpoints are then implemented and kept unchanged until the next dispatch period $t+1$.

\subsection{Control of PV Inverters}
Encouraged by interconnection standards (e.g., IEEE 1547) and possible incentive schemes (e.g., \cite{Incentive}), inverters are starting to play a more active and important role in system operation. Meanwhile, various standards have been developed to guide their interoperability and communication (e.g., IEEE 1815, IEEE 2030.5, IEC 61850), which makes it possible for the smart inverters to solve the lower-level optimization problem autonomously and cooperatively. Although individual inverters usually do not have direct access to global information about the system, they are able to take rapid control actions by measuring their local voltage magnitude and exchanging information with other inverters via Wi-Fi, ZigBee or power line communication (PLC). These capabilities allow them to iteratively adjust their output towards the satisfaction of the KKT conditions by implementing fully-distributed control strategies \cite{Communication, Dualascent1, Dualascent2, Primaldual1}. These strategies are of different design, for example, the strategy proposed in \cite{Communication} adds an additional feedback loop to deal with the reactive power constraint, which is handled by a special quadratic penalty function in \cite{Primaldual1}. \cite{Dualascent1} focuses on the limited bandwidth of the communication link, whereas \cite{Dualascent2} addresses the delayed communication. However, the execution of these strategies are similar. As Fig.~\ref{Framework} shows, at each iteration cycle $k \in [t, t+1]$, an inverter performs the following four major steps:
\begin{itemize}
    \item Measure the local voltage magnitude; 
    \item Calculate the control signal based on a pre-defined logic; 
    \item Implement the control signal to adjust its reactive output; 
    \item Exchange information with neighbors. 
\end{itemize}
These four steps are similar to one iteration step in the dual ascent or primal-dual gradient algorithms. As these algorithms solve the lower-level optimization problem iteratively, the inverter output converges asymptotically to the optimal value. The details of the strategies and the theoretical proof of convergence and optimality can be found in \cite{Communication, Dualascent1, Dualascent2, Primaldual1}. 

Moreover, the performance of the feedback optimization-based strategy, such as its robustness against measurement noise, real-time computational feasibility, and effectiveness under asynchronous communication have been validated by field test results \cite{Validation}. The practicality of organizing smart inverters under a multi-agent peer-to-peer communication framework for distribution network voltage regulation has also been demonstrated in real-world project by the Toronto and Region Conservation Authority (TRCA)\cite{Validation1}.

\subsection{Discussion}
The ADMS uses feedforward optimization process. It models the coordinated actions of the PV inverters based on its full knowledge of the network and is thus able to influence those actions indirectly by changing the PV inverters' local voltage magnitude using the OLTCs and CBs. In this way, the ADMS exploits the reactive power compensation capability of the PV inverters for system-wide optimal VVC while respecting their autonomy. On the other hand, the feedback optimization-based lower-level solution approach enables the PV inverters to take actions based on the constant monitoring of their local voltage. The switching of the mechanical devices are essentially taken into consideration when they optimize their goals.

\section{Case Studies}
In this section, we evaluate the performance of the proposed approach in different systems under various scenarios. All case studies were performed using MATLAB 2018a on a laptop with 2.2 GHz Intel Core i5-10310 CPU and 16 GB RAM. The nonlinear AC power flow model of the test systems were simulated using MATPOWER \cite{MATPOWER}. The optimization models were solved using Gurobi \cite{Gurobi} with CVX \cite{CVX}.

\subsection{Test Systems}
\begin{figure}[htbp!]
\centering

\begin{subfigure}[b]{0.5\textwidth}
   \includegraphics[width=3.5in]{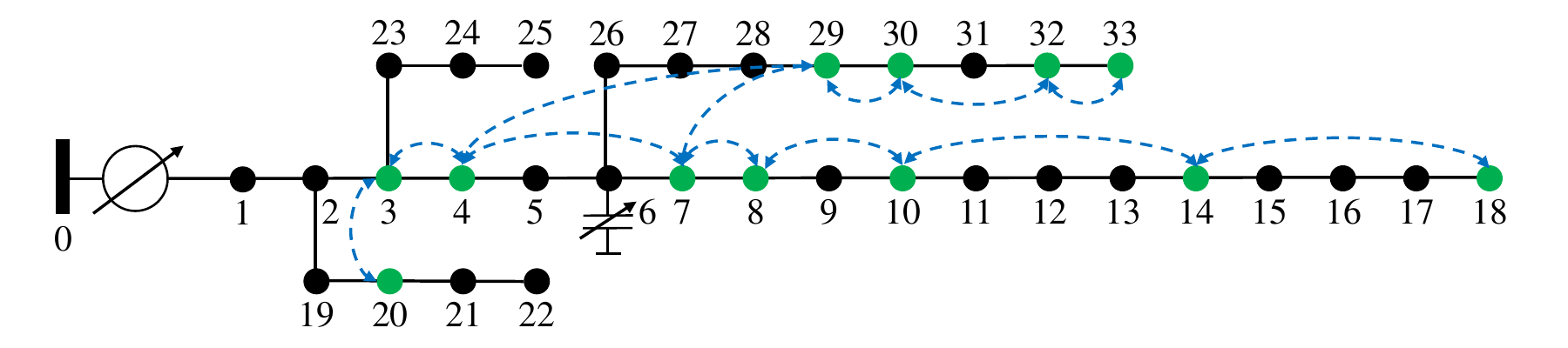}
   \caption{}
\end{subfigure}

\begin{subfigure}[b]{0.5\textwidth}
   \includegraphics[width=3.5in]{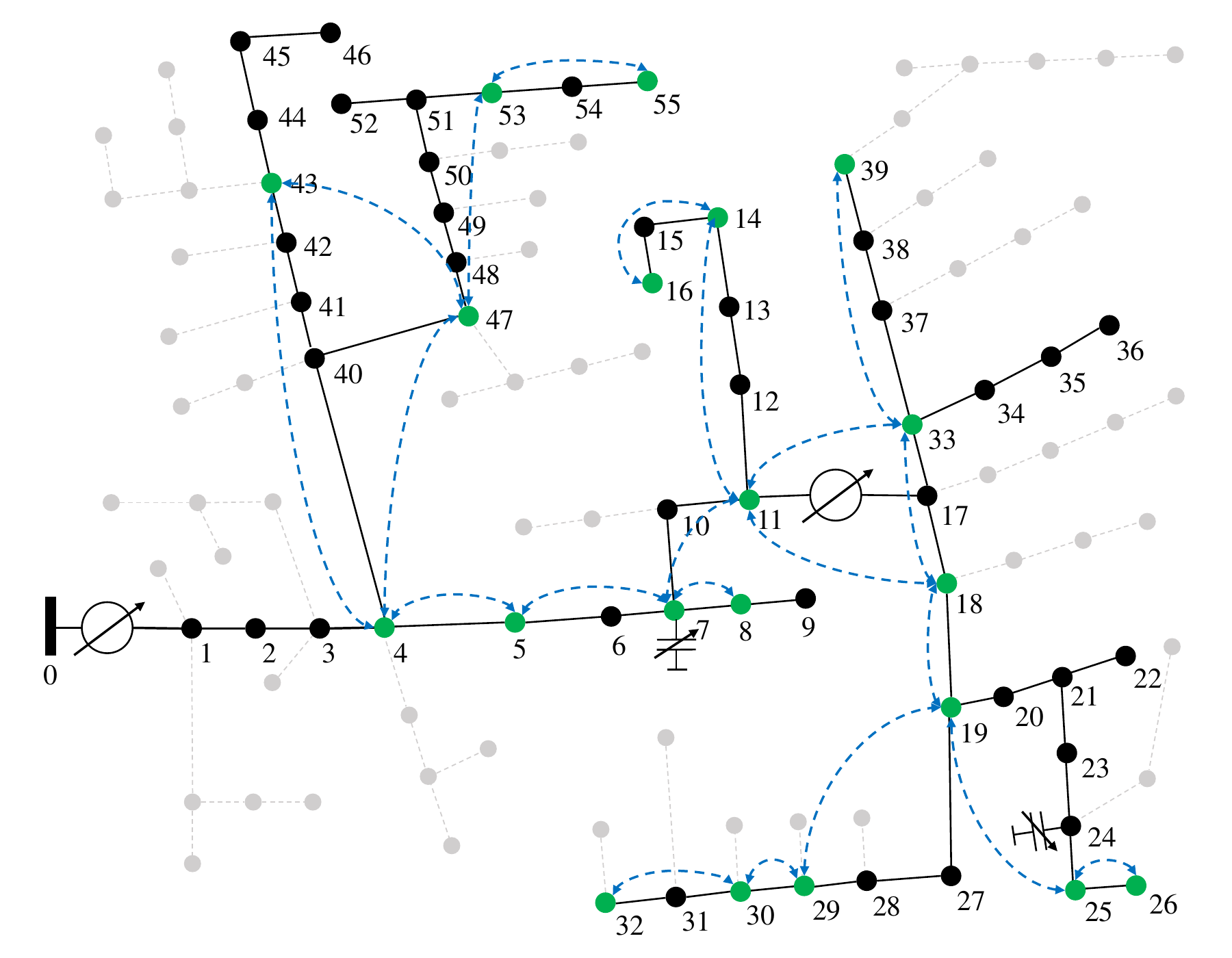}
   \caption{}
\end{subfigure}

\caption{Single-line diagram of the test system (a) 33-bus network; (b) 123-bus network.}
\label{Network}
\end{figure}

\begin{figure}[ht!]
\centering
\includegraphics[width=3.45in]{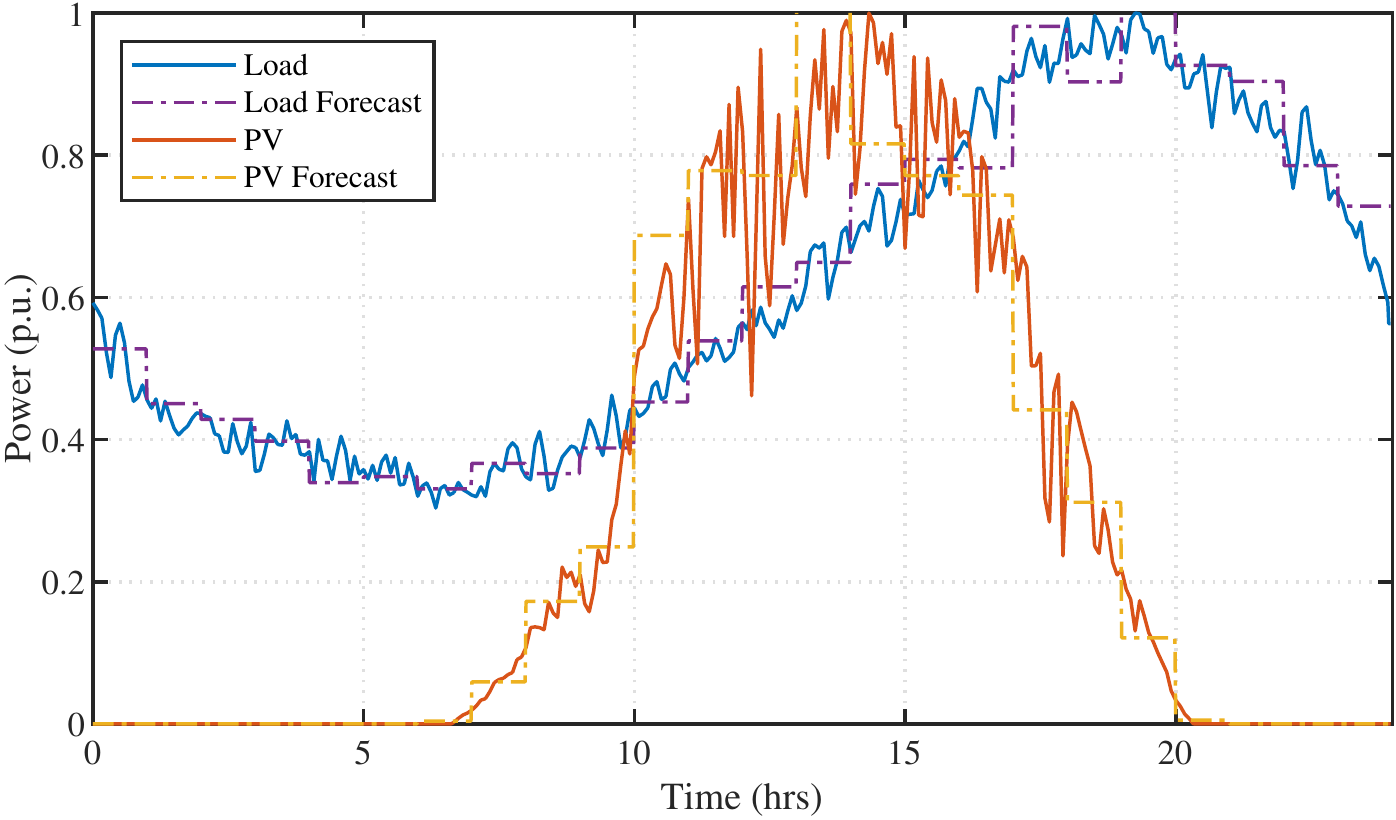}
\caption{Normalized daily profiles and forecast of load and PV.}
\label{LPV}
\end{figure}

Fig.~\ref{Network} shows the single-line diagram of the balanced IEEE 33-bus radial distribution network \cite{DistFlow} and the three-phase backbone of the IEEE 123-bus distribution test feeder \cite{Communication} that are used to demonstrate the performance of the proposed VVC framework. The black and green dots represent the load and PV nodes. The blue dashed lines are the communication links between neighboring PV inverters. Two inverters are defined as neighboring inverters if the path that connects them does not pass any other inverters \cite{Communication}. Table~\ref{tab: PV capacity 33} and \ref{tab: PV capacity 123} show the capacities of the aggregated residential PV systems distributed in the test systems. The 1-minute-resolution daily profiles for PV generation and load consumption was constructed based on the Pecan Street data set of June 16, 2014 \cite{Pecan}. The hourly PV and load forecast value used by the ADMS are their hourly average value with a random percent error, $\pm 20\%$ for PV generation and $\pm 10\%$ for load consumption. Fig.~\ref{LPV} shows their normalized profiles and forecast values over the course of one day. Each PV inverter is oversized by $10\%$ so it can provide reactive power compensation while generating at its rated active power capacity \cite{1547}. Each OLTC allows for $\pm 0.05$ p.u. voltage regulation. Each of its 17 tap positions, labeled as ${\{-8,-7,...,-1,0,1,...,7,8\}}$, provides 0.00625 p.u. regulation. The maximum number of OLTC tap changes between two dispatch periods is 3. Every CB includes 3 switchable units and each unit is rated at 100 kVAr. The number of units switched ON corresponds to the status of the CB, i.e, ${\{0,1,2,3\}}$. Only one unit change in the number of connected CB units is allowed between two dispatch periods. The acceptable voltage range is 0.95 p.u. to 1.05 p.u. The upper-level sends hourly setpoint to the mechanical devices while the PV inverters in the lower-level adjust their outputs every 0.5 seconds.

\begin{table}[htbp!]
\renewcommand{\arraystretch}{1.15}
    \centering
    \caption{\textsc{Capacities of the PV Systems in 33-bus System}}
    \label{tab: PV capacity 33}
\begin{tabular}{c c c c c c}
    \toprule
    Capacity (kW) & 200  & 300  & 400  & 600\\
    \midrule
    Node & 3, 7, 20, 29, 32  & 4, 8, 30 & 10, 33 & 14, 18\\
    \bottomrule
\end{tabular}
\end{table}

\begin{table}[htbp!]
\renewcommand{\arraystretch}{1.15}
    \centering
    \caption{\textsc{Capacities of the PV Systems in 123-bus System}}
    \label{tab: PV capacity 123}
\begin{tabular}{c c c c c c}
    \toprule
    Capacity (kW) & 50  & 100  & 200  & 300\\
    \midrule
     & & 5, 7, 8, &16, &\\
    Node & 4, 11, 14  &  18, 19, 29, 30,  &25, 39, & 26, 32, 55\\
    & & 33, 43, 47 &53 &\\
    \bottomrule
\end{tabular}
\end{table}

\subsection{Performance Analysis}
This subsection illustrates the operation of the proposed approach through a 24-h simulation of the IEEE 33-bus network. The PV inverters in the lower-level implement the strategy proposed in \cite{Primaldual1} to adjust their outputs based on the objective function ${f(\textbf{q}_t^g) = a_i^2(q_{i,t}^g)^2 + ({\textbf{q}_t^g})^TX_{gg}(\textbf{q}_t^g)}$, where $a_i$ is a synthetic value in the range of 0.5--1.5.

\subsubsection{Exactness and Convergence}
\begin{figure}[ht!]
\centering
\includegraphics[width=3.45in]{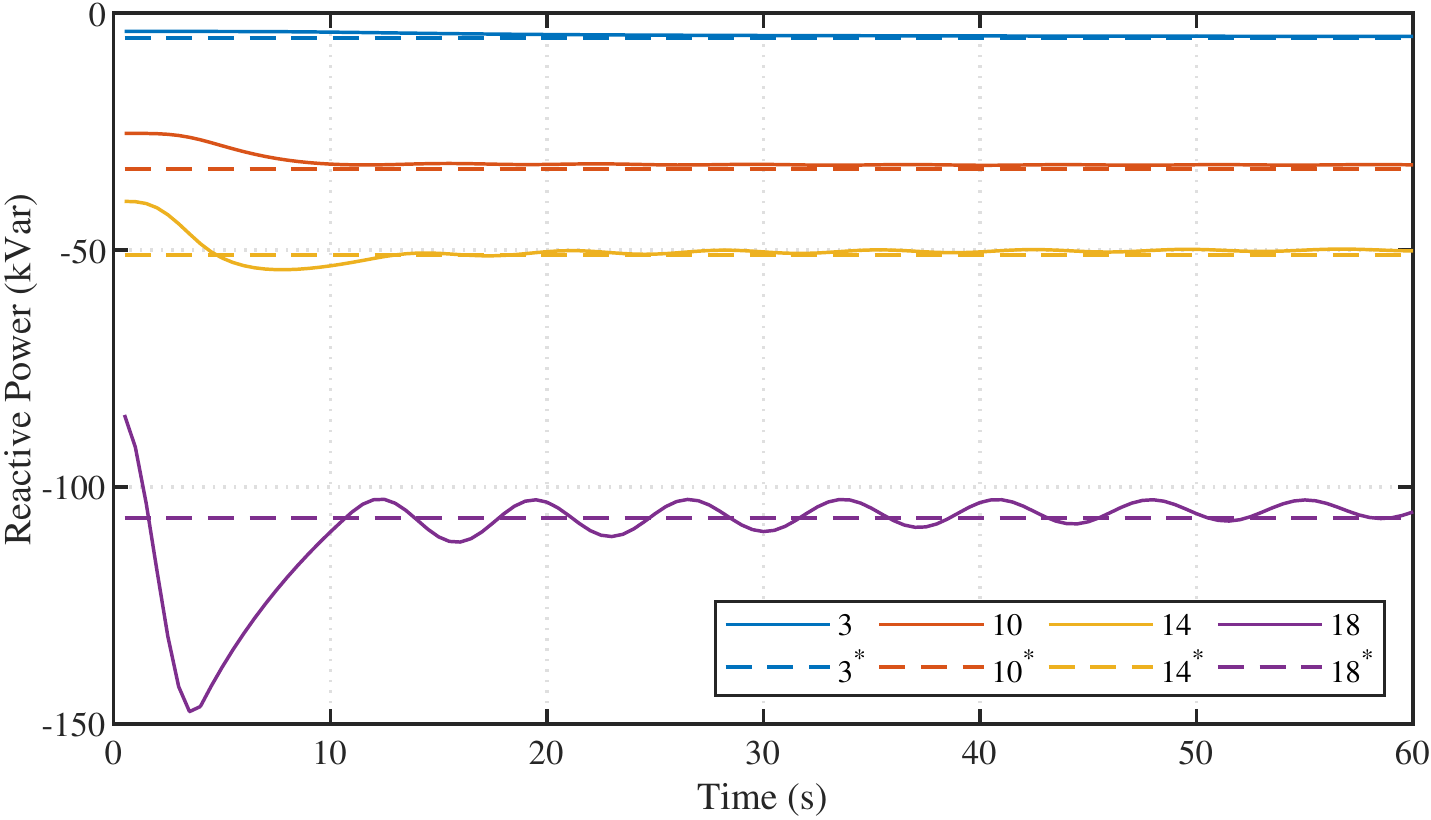}
\caption{PV inverter reactive output iteration process at 13:00.}
\label{COPT}
\end{figure}

Over the course of the day, the MISOCP model in (\ref{eq: fo}) is solved 24 times to determine the hourly settings of the mechanical devices. The average solution time is 1.70 seconds. In addition, the maximum conic relaxation gap $\epsilon = \sum_{(i, j) \in \mathcal{E}}|l_{ij,t} - (P_{ij,t}^2+Q_{ij, t}^2)/v_{i,t}|$ is $1.30\times10^{-5}$, which shows that the conic relaxation is sufficiently close to the original non-convex model. The PV inverters in the lower-level adjust their reactive outputs based on their real-time local voltage and neighboring communication. Fig.~\ref{COPT} shows the iteration process of their reactive outputs at 13:00. After about 15 seconds, those outputs (solid lines) have converged to the optimal solution of problem (\ref{eq: lower level}) (dashed lines). That is, the PV inverters are able to cooperate to drive the system to the satisfaction of the KKT conditions of problem (\ref{eq: lower level}), which validates the accuracy of the way the bi-level optimization model captures the operation characteristics of the coordinated PV inverter group.

\subsubsection{Control Results Over a 24-h Period}
\begin{figure}[ht!]
\centering
\includegraphics[width=3.45in]{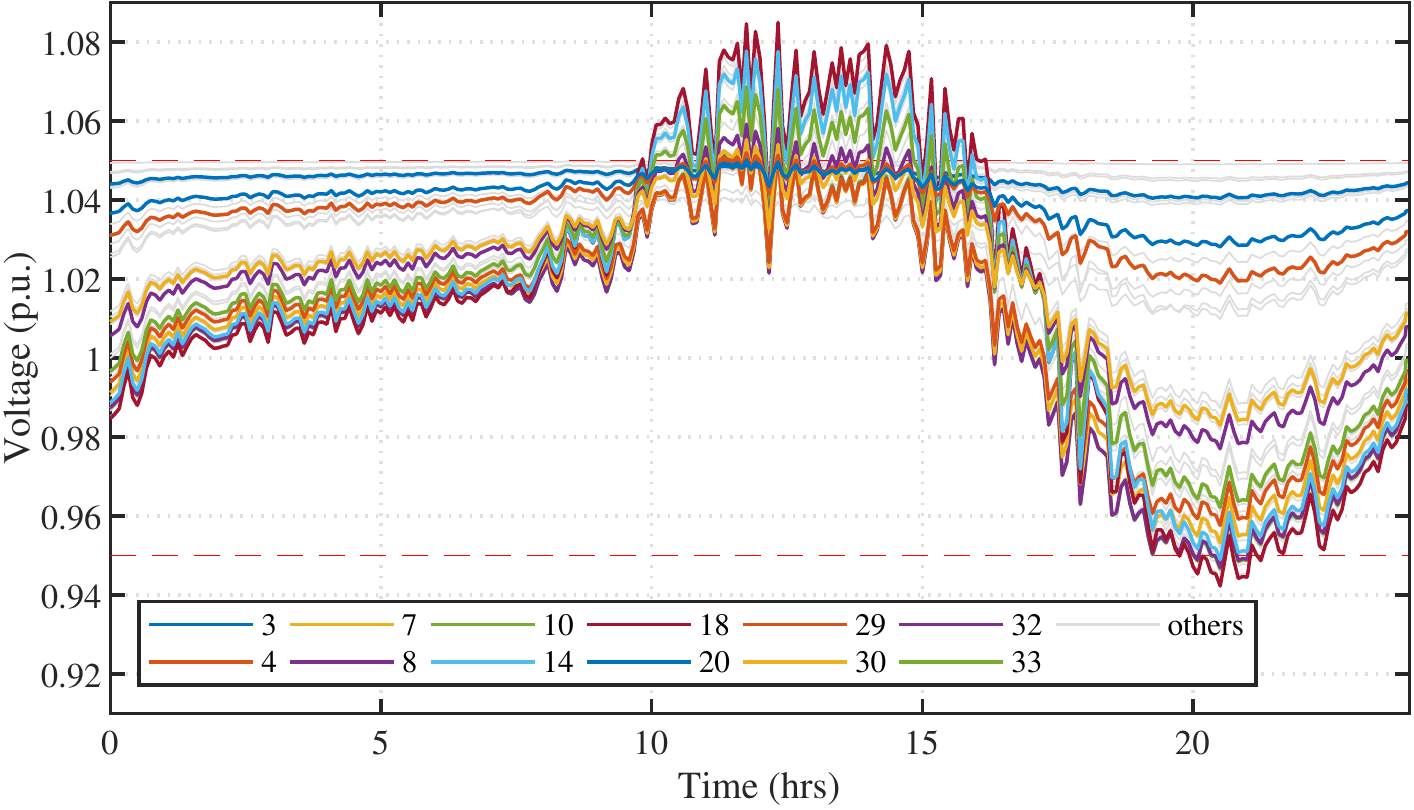}
\caption{Network voltage profiles without any control.}
\label{NonConVol}
\end{figure}

\begin{figure}[ht!]
\centering
\includegraphics[width=3.45in]{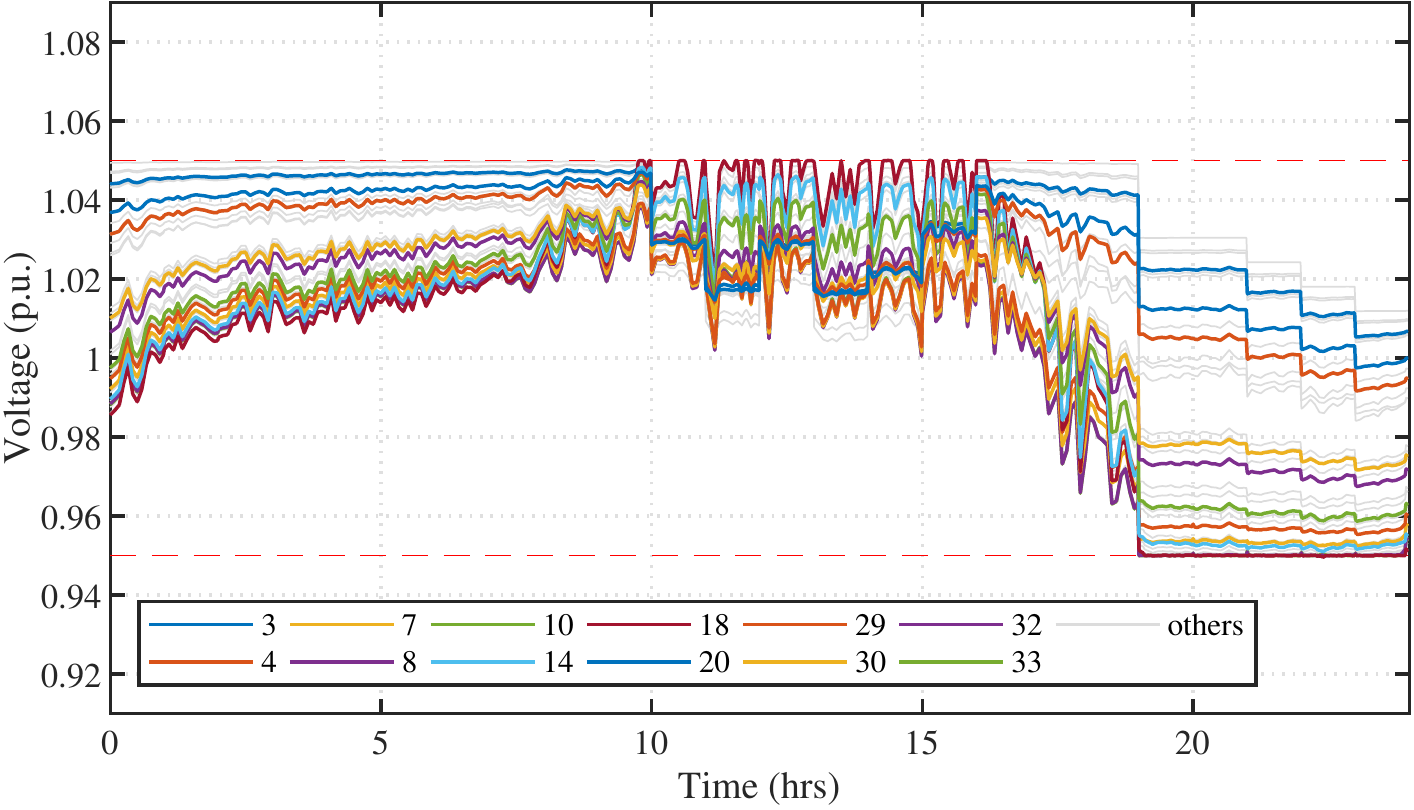}
\caption{Network voltage profiles with the proposed control.}
\label{ConVol}
\end{figure}

\begin{figure}[ht!]
\centering
\includegraphics[width=3.45in]{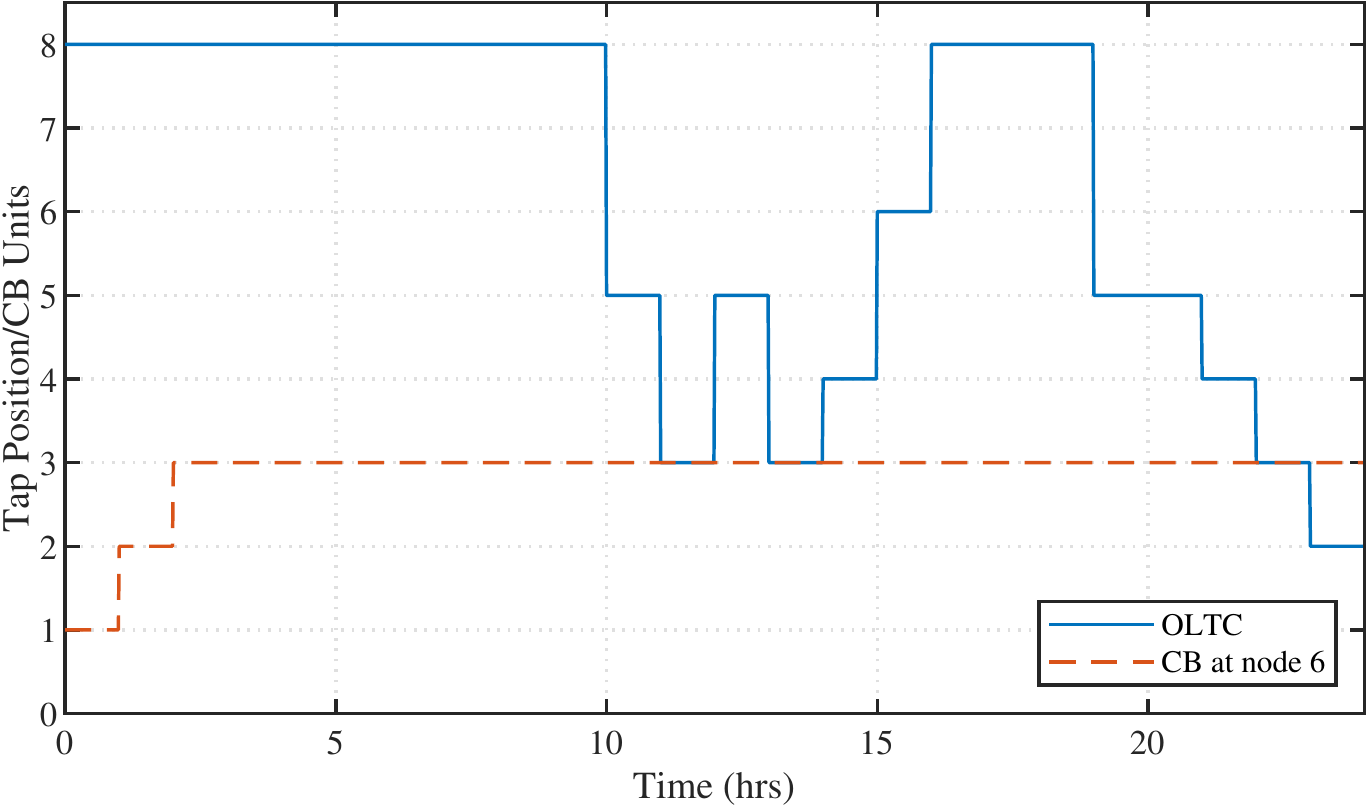}
\caption{Switching actions of the mechanical devices.}
\label{TCB}
\end{figure}

\begin{figure}[ht!]
\centering
\includegraphics[width=3.45in]{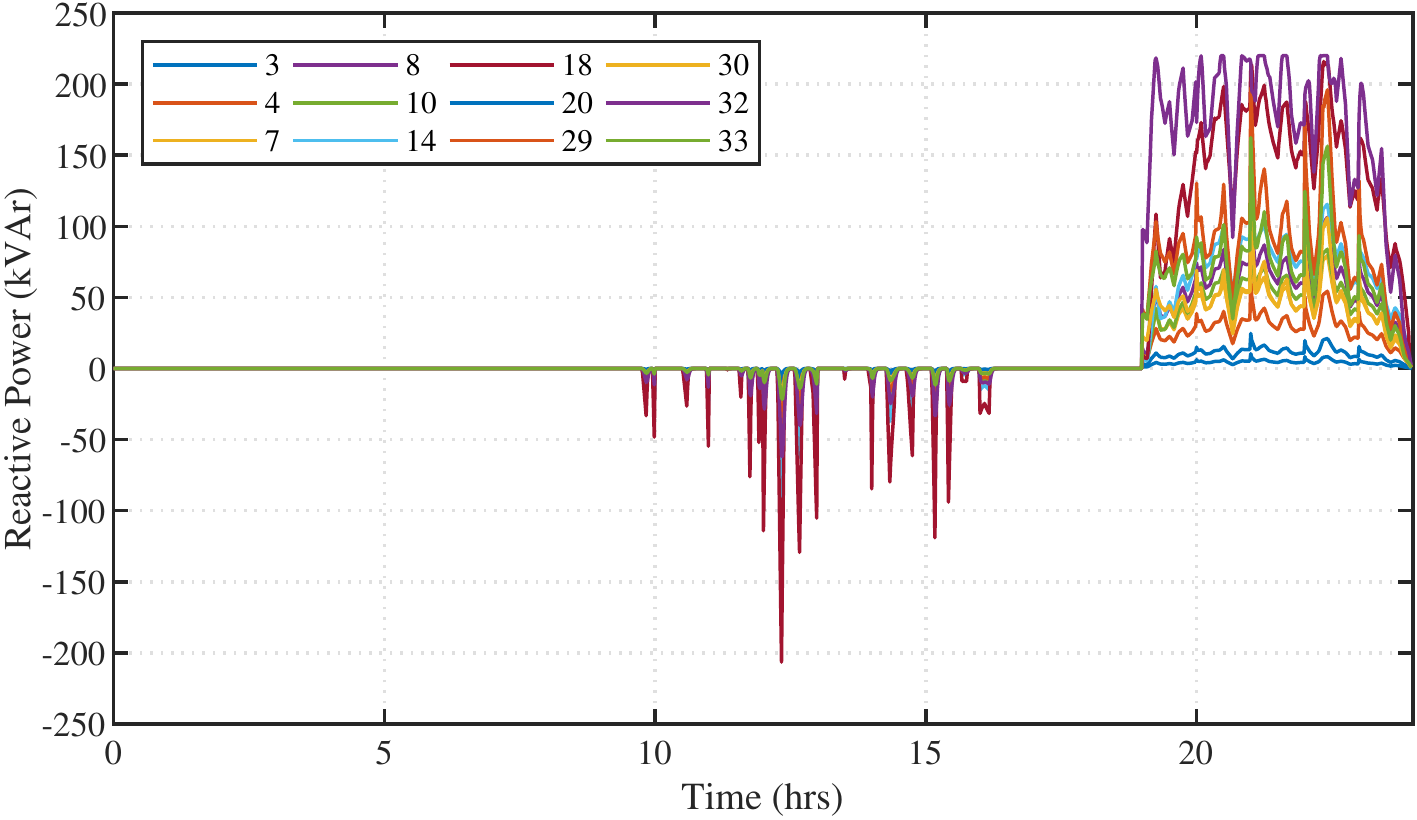}
\caption{PV inverter reactive power outputs.}
\label{PVq}
\end{figure}

Fig.~\ref{NonConVol} shows the voltage profiles of the IEEE 33-bus network when no control strategy is implemented. Due to the fast fluctuations in load and PV generation, the voltage magnitude changes rapidly. Moreover, the system experiences over-voltage issues at midday, when the high PV generation exceeds the local load and causes a reversal in the power flow. On the other hand, after 19:00, the heavy load leads to under-voltage problems. Fig.~\ref{ConVol}-\ref{PVq} show how the proposed approach is able to coordinate all the voltage regulation devices to effectively constrain the voltage profiles within the specified range. Before 19:00, the ADMS adjusts the OLTC tap position and the number of connected CB units to maintain a relatively high voltage level across the network so as to reduce the active power losses. During this period, since the ADMS only updates the setpoints of the OLTC and CBs on an hourly basis and forecast errors arise, voltage limit violations occur within the dispatch period. When the fast-reacting PV inverters detect these violations, they cooperate to remove them at the minimal value of their group objective. After 19:00, however, the ADMS determines that it can minimize the network active power losses by forcing the PV inverters to provide reactive power to mitigate the loading of the distribution lines. Therefore, the ADMS switches down the mechanical devices and deliberately sets the voltage magnitude of node 18 at the lower voltage limit. As a result, the PV inverter group provides a consistently high amount of reactive power to the network, which alleviates the loading of the distribution network and reduces the active power losses. This reactive power varies in response to the fast fluctuations in the load and the switching of the mechanical devices. The proposed approach effectively reduces the average daily active power loss $\bar{P}_{\text{loss}}=\frac{1}{\sum t}\sum_t\sum_{(i, j) \in \mathcal{E}}r_{ij}l_{ij, t}$ across the network by $15.7\%$, from 102.44 kW to 86.34 kW.

\subsection{Comparison with Other Approaches}
This section compares the performance of the proposed approach with other optimization models. In this study, the PV inverters in the lower-level cooperate as described in subsection IV-B. The OLTC and CBs are dispatched by the proposed bi-level optimization model as well as two other traditional single-level optimization models. These single-level optimization models are similar to the upper-level MISOCP model, while the PV inverter reactive power outputs are treated in two different ways: 
\begin{itemize}
    \item Model 1: Assume the PV inverters follow the centrally-dispatched setpoints and directly include $\textbf{q}_t^g$ as an upper-level decision variable. This is one of the most common ways of modeling the PV inverter reactive output and has been adopted in many multi-level VVC approaches, e.g. \cite{Twolevel, ADMM1, ADMM2, Stochastic, Robust};
    \item Model 2: Ignore the PV inverter reactive power output and remove $\textbf{q}_t^g$ from the optimization model. Since the PV inverters no longer follow the dispatched setpoints, their outputs are not directly controllable by the ADMS. One natural way is to treat the PV nodes as simple residential load nodes and ignore the reactive output of the PV inverters.
\end{itemize}

Table~\ref{tab: 33 Results} and \ref{tab: 123 Results} compare these three models' exactness (i.e., maximum conic relaxation gap $\epsilon$), complexity (i.e., average solution time $t_s$) as well as optimality (i.e., average daily network active power loss $\bar{P}_{\text{loss}}$) in the IEEE 33-bus and 123-bus systems. In all these cases, the conic relaxation is close to the original non-convex model. The bi-level model and Model 1 take longer to solve as they involve more optimization variables compared to Model 2. Moreover, as the bi-level model introduces extra auxiliary variables, its solution time is the longest and this time increases faster with the size of the problem. On the other hand, the bi-level approach effectively models the PV inverter group operation characteristic, and thus achieves much smaller average daily active power loss. Model 1 is overly optimistic about the participation of the consumer-owned PV inverters in the optimization of the utility's VVC objective, while Model 2 is unable to harness the reactive output from the PV inverters for minimizing network active power losses. As a result, $\bar{P}_\text{loss}$ with Model 1 and Model 2 are $12.0\%$ and $8.4\%$ higher than that with the proposed bi-level optimization model in the 33-bus test case. Similar results can be found in the 123-bus test case.

\begin{table}
\renewcommand{\arraystretch}{1.15}
    \centering
    \caption{\textsc{Comparison Results in 33-bus System}}
    \label{tab: 33 Results}
\begin{tabular}{c c c c}
    \toprule 
     & Bi-level  & Model 1  & Model 2 \\
    \midrule
    Maximum relaxation gap $\epsilon$ ($\times10^{-5})$ & 1.30 & 1.24  & 0.21 \\
    \midrule
    Average solution time $t_s$ (s) & 1.70  & 1.64  & 1.41 \\
    \midrule
    Average power loss $\bar{P}_\text{loss}$ (kW) & 86.34  & 96.68  & 93.59 \\
    \bottomrule
\end{tabular}
\end{table}

\begin{table}
\renewcommand{\arraystretch}{1.15}
    \centering
    \caption{\textsc{Comparison Results in 123-bus System}}
    \label{tab: 123 Results}
\begin{tabular}{c c c c}
    \toprule 
     & Bi-level  & Model 1  & Model 2 \\
    \midrule
    Maximum relaxation gap $\epsilon$ ($\times10^{-5})$ & 0.47  & 0.29 & 0.44 \\
    \midrule
    Average solution time $t_s$ (s) & 5.48  & 2.31 & 2.00 \\
    \midrule
    Average power loss $\bar{P}_\text{loss}$ (kW) & 73.09  & 88.04  & 77.21 \\
    \bottomrule
\end{tabular}
\end{table}

\subsection{Generalizability Test}
This subsection evaluates the performance of the proposed approach in the IEEE 33-bus system when the upper-level or lower-level model is different from those implemented in Section IV-B simulation case.
\subsubsection{Upper-level MILP Optimization Model}
Besides the MISOCP model, MILP is another widely adopted optimization technique for dispatching the mechanical devices. The column labeled ``MILP'' in Table~\ref{tab: General} shows that when the upper-level model is replaced by the MILP model, the solution time decreases significantly while the average daily network active power loss remains close. This is mainly because although the MILP model simplifies the power flow model and thus results in different mechanical device setpoints, the PV inverter group would adjust their outputs correspondingly to minimize the network active power losses.
\subsubsection{Lower-level Equalization of the Utilization Ratios}
The column labeled ``Equal'' in Table~\ref{tab: General} shows the simulation results when the PV inverters in the lower-level implement a consensus algorithm \cite{Consensus2,Consensus3} to adjust their outputs for achieving equal utilization ratios. Since the lower-level objective no longer includes minimizing the network active power losses, the average daily active power loss is slightly higher. The average solution time is approximately the same.
\subsubsection{Lower-level with Several Inverter Groups}
The column labeled ``Group'' in Table~\ref{tab: General} shows the simulation results when the 33-bus system is partitioned into 3 sub-networks \cite{Group} and thus contains 3 separate PV inverter groups: \{3, 4, 7, 8, 20\}, \{10, 14, 18\}, \{29, 30, 32, 33\}. In this case, as the cooperation scope of the PV inverters is limited, the average daily active power loss is also slightly higher. The average solution time is barely impacted.

Overall, the proposed bi-level optimization model can be flexibly adapted to incorporate different upper-level and lower-level optimization models.

\begin{table}
\renewcommand{\arraystretch}{1.15}
    \centering
    \caption{\textsc{Performance with Varying Models}}
    \label{tab: General}
\begin{tabular}{c c c c}
    \toprule 
     & MILP  & Equal  & Group \\
    \midrule
    Maximum relaxation gap $\epsilon$ ($\times10^{-5})$ & - & 0.59  & 0.69 \\
    \midrule
    Average solution time $t_s$ (s) & 0.87 & 1.55 & 1.64 \\
    \midrule
    Average power loss $\bar{P}_\text{loss}$ (kW) & 84.86  & 91.06  & 88.32 \\
    \bottomrule
\end{tabular}
\end{table}

\section{Conclusion}
This paper proposes a bi-level optimization-based VVC framework for distribution networks with both conventional voltage control devices and smart residential PV inverters. The centralized upper-level optimization determines the setpoints for the OLTCs and CBs on an hourly-basis by solving a MISOCP optimization problem based on the single-level reduction approach. This level takes advantages of the network information available to the ADMS and achieves system-wide optimal coordination among all control devices. The distributed lower-level optimization exploits the autonomous capability of the smart PV inverters by enabling them to adjust their outputs based on local voltage measurements and neighboring communication. It corrects the sudden voltage violations that occur during the upper-level dispatch periods and also achieves an optimal utilization of all the available inverter reactive power capacity. This VVC framework successfully captures the interactions between the mechanical devices and the autonomous inverters. Hence, it facilitates the coordination between these devices and fully exploits their capabilities for voltage regulation. This framework represents an effective and flexible way for the distribution networks to accommodate autonomous PV inverters in the VVC architecture. Future research topics include extending the proposed approach to three-phase unbalanced networks, where the voltage imbalance problem can be further addressed, accommodating other types of voltage regulation resources, e.g., PV inverter active power, and analyzing the exactness of the bi-level SOCP model theoretically.

\ifCLASSOPTIONcaptionsoff
  \newpage
\fi

\appendices
\section{Proof of (\ref{eq: cons})}
The lower-level optimization model for equalizing the utilization ratios can be summarized as follows (the subscript $t$ is omitted in this section):
\begin{equation} \label{eq: conopt}
\begin{split}
\min_{\textbf{q}^g} \quad\quad&\sum_{i \in \mathcal{N_G}} \frac{X_{li}}{2\bar{q}_{i}}(q_{i}^g)^2 \\
\text{s.t.} \quad\quad&\ubar{v} \le v_l^g \le \bar{v},   \\
&\ubar{q}_i  \le q_{i}^g \le \bar{q}_i, \: \forall i \in \mathcal{N_G}
\end{split}
\end{equation}
where $v_l^g$ is the local voltage of the leader inverter. The voltage constraint ${\ubar{v} \le v_{i} \le \bar{v}}$, $\forall i \in \mathcal{N_G}$ in (\ref{eq: lower level}) is simplified as ${\ubar{v} \le v_{l}^g \le \bar{v}}$ here because the leader inverter is assumed to experience the largest voltage deviation. Assume problem~(\ref{eq: conopt}) is feasible and the Slater condition is satisfied, i.e. there exist ${\ubar{q}_i < q_{i}^g < \bar{q}_i}$, ${\forall{i \in \mathcal{N_G}}}$ such that ${\ubar{v} < v_l^g < \bar{v}}$, given the strong convexity of the cost function, this problem has a unique optimal solution. 

To prove that this optimal solution promotes an equal utilization ratio in the PV inverter group, we first ignore the reactive power constraint and obtain the following Lagrangian function:
\begin{equation} \label{eq: lac}
    \mathcal{L}(q_{i}^{g}, \ubar{\xi}, \bar{\xi}) = \sum_{i \in \mathcal{N_G}} \frac{X_{li}}{2\bar{q}_{i}}(q_{i}^g)^2+\ubar{\xi}(\ubar{v}-v_l^g)+\bar{\xi}(v_l^g-\bar{v})
\end{equation}
where $\ubar{\xi}$ and $\bar{\xi}$ are the dual variables for the lower and upper voltage limit constraints. The derivative of  ${\mathcal{L}(q_{i}^{g}, \ubar{\xi}, \bar{\xi})}$ is:
\begin{equation} \label{eq: dlac}
    \nabla_{q_{i}^{g}}\mathcal{L}(q_{i}^{g}, \ubar{\xi}, \bar{\xi}) = \frac{X_{li}}{\bar{q}_{i}}q_{i}^g+(\bar{\xi}-\ubar{\xi})\frac{\partial v_l^g}{\partial q_i^g}
\end{equation}
where ${\partial{v_l^g}/\partial{q_i^g}}$ is the sensitivity of the leader inverter's local voltage to inverter $i$'s reactive output. By approximating it as $X_{li}$ \cite{Validation} and setting ${\nabla_{q_{i}^{g}}\mathcal{L}(q_{i}^{g}, \ubar{\xi}, \bar{\xi})=0}$, we obtain the critical point:
\begin{equation} \label{eq: criticalpoint}
    q_{i, \text{unc}}^g = \bar{q}_{i}(\ubar{\xi}-\bar{\xi}).
\end{equation}
This is the critical point when the reactive output is unconstrained. The solution to the constrained optimization problem~(\ref{eq: conopt}) is obtained by projecting $\textbf{q}_{\text{unc}}^g$ onto the set of feasible reactive output region. Due to the special structure of the objective function, this can be handled by projecting each $q_{i, \text{unc}}^g$, ${\forall i \in \mathcal{N_G}}$ to its independent bounds ${[\ubar{q}_i, \: \bar{q}_i]}$. Details of the proof can be found in \cite{Dualascent2}. Thus, we have the constrained optimizer: 
\begin{equation} \label{eq: qopt}
    q_{i, \text{opt}}^g = \left[ \bar{q}_{i}(\ubar{\xi}-\bar{\xi})\right]_{\underline{q}_i}^{\bar{q}_i}.
\end{equation}
The utilization ratio of the inverter is:
\begin{equation} \label{eq: uopt}
    u_i = \frac{q_{i, \text{opt}}^g}{\bar{q}_i} = \left[(\ubar{\xi}-\bar{\xi}) \right]_{-1}^{1}.
\end{equation}
That is, a group of PV inverters that cooperate to solve the optimization problem~(\ref{eq: conopt}) will converge on the same utilization ratio. 

% \begin{IEEEbiography}[
%     {\includegraphics[width=1in,height=1.25in,clip,keepaspectratio]{Yao.jpg}}
%     ]{Yao Long}
%     (S'17) 
%     received the B.Eng. degree in Electrical Engineering and Automation from Huazhong University of Science and Technology, Wuhan, China, in 2014. She obtained her M.S. and Ph.D.\ degrees in Electrical Engineering from the University of Washington, Seattle, WA, USA in 2019 and 2021, respectively. Her research interests include power system control and optimization, smart inverter and renewable generation.
% \end{IEEEbiography}

% \begin{IEEEbiography}[
%     {\includegraphics[width=1in,height=1.25in,clip,keepaspectratio]{Daniel.jpg}}
%     ]{Daniel S.\ Kirschen}
%     (M’86–SM’91–F’07)
%     is the Donald W. and Ruth Mary Close Professor of Electrical and Computer Engineering at the University of Washington. His research focuses on the integration of renewable energy sources in the grid, power system economics and power system resilience. Prior to joining the University of Washington, he taught for 16 years at The University of Manchester (UK). Before becoming an academic, he worked for Control Data and Siemens on the development of application software for utility control centers. He holds a PhD from the University of Wisconsin-Madison and an Electro-Mechanical Engineering degree from the Free University of Brussels (Belgium). He is the author of two books and over 200 scientific papers. He is a Fellow of the IEEE.
% \end{IEEEbiography}

\end{document}